\documentclass[11pt]{article}

\usepackage[margin=0.9in]{geometry}
\usepackage{amsmath,amssymb,amsthm,mathtools,bm}
\usepackage{booktabs}
\usepackage{microtype}
\usepackage{xcolor}
\usepackage[hidelinks]{hyperref}
\usepackage[numbers,sort&compress]{natbib}
\newtheorem{theorem}{Theorem}
\newtheorem{proposition}{Proposition}

\theoremstyle{remark}
\newtheorem{remark}{Remark}
\newcommand{\Ad}{\operatorname{Ad}}
\newcommand{\Tr}{\operatorname{Tr}}

\newcommand{\I}{\mathcal I}
\newcommand{\C}{\mathsf C}
\newcommand{\R}{\mathsf R}
\newcommand{\D}{\mathsf D}
\newcommand{\G}{\mathcal G}
\newcommand{\ket}[1]{\lvert #1\rangle}
\newcommand{\bra}[1]{\langle #1\rvert}

\title{Reference-Assisted Qutrit Control and Conditional One-Round Distillation in a Charge-4e Superconductor Model}
\author{Junhao (Eigen) Wang\\[2pt] \small Department of Mathematics, New York University, New York, NY, USA\\ \small \texttt{jw10111@nyu.edu}}
\date{November 2026\\[2pt]\small Preprint v0.6 (November 19)}

\begin{document}
\maketitle

\begin{abstract}
We formulate a charge-$4e$ parafermionic-vortex route to qutrit processing as a conditional theory with explicit operational assumptions. Sector preservation and same-label charge-conjugation covariance imply closure for every adaptive visible record. For charge-conjugation-invariant inputs, record probabilities are independent of the supported reference state; for arbitrary inputs, discarding the reference gives a population-weighted mixture of conjugate sector maps. We give a finite-depth stability bound for complete labelled instruments, including postselection and initial zero-sector contamination. A self-contained calculation reproduces the known dense single-qutrit closure of the registered Clifford-plus-Flip gate set. We also specify an abstract ternary CSS $[[11,1,5]]_3$ syndrome/recovery decoder: 243 classical representatives per CSS component correct all 3,609 Pauli errors of support weight at most two, and an exact integer dynamic program yields the full independent-Pauli logical enumerator. Its all-syndrome acceptance is distinct from the known $1/1728$ ideal acceptance of one-round strange-state Golay distillation. An accepted-run target-infidelity budget and a synthetic parameter protocol expose the remaining noise assumptions. Native charge-$4e$ syndrome extraction, material feasibility, a device threshold, and unconditional fault tolerance are not established.
\end{abstract}

\section{Introduction}

A charge-$4e$ topological superconductor with chiral $\mathbb Z_3$ topological order, described by $U(2)_{4,0}$ Chern--Simons theory, has been proposed as a platform in which flux defects carry parafermionic zero modes and can implement a qutrit version of topological quantum processing \cite{Shi2026}.  The attractive division of labor is familiar: defect braiding supplies protected algebraic operations, while a coherent interferometric measurement supplies a non-Clifford resource.  Turning this division into a logically closed protocol requires more than checking individual braid matrices.  One must identify the source of the Pauli operators, specify what an external data qutrit experiences when the charge-conjugation sector is not observed, and place every non-topological step inside a quantitative error model.  The broader background already includes general magic-state routes \cite{BravyiKitaev2005}, ternary-Golay distillation \cite{Prakash2020}, and measurement-assisted universality in other anyon models \cite{CuiHongWang2014,Chen2025}; our claims are therefore limited to the reference-assisted instrument formulation and error-interface analysis below. The gate-density and Golay-distillation results are reproduced background, not new universality or distillation protocols.

Higher-charge superconductivity supports distinct topological orders at the same condensate charge. Recent work has developed a clustering hierarchy of topological charge-$2ne$ superconductors~\cite{Gao2026Clustering}, charge-$6e$ constructions from doped $SU(3)$ spin liquids~\cite{Wang2026Charge6e}, and a classification based on how superconductivity and intrinsic topological order intertwine~\cite{Gao2026SCSharp}. In particular, the $U(2n)_{2n,0}$ family includes a charge-$4e$ phase described by $U(4)_{4,0}$, distinct from the $U(2)_{4,0}$ phase underlying the $\mathbb Z_3$ defect model considered here. The reference-assisted instrument results below concern the specified operations in the latter model; they do not classify the computational power of vortices in other higher-charge superconductors.

This paper isolates those questions in a single idealized operation model.  Our first contribution is an explicit reference-assisted instrument formulation, related to the general reference-frame and superselection framework~\cite{Bartlett2007}. Its adaptive closure theorem specializes these symmetry ideas to the stated visible-record conventions.  A data qutrit is coupled to a reference qutrit whose usable state lies in the two-dimensional sector span $\operatorname{span}\{\ket{1},\ket{2}\}$.  The two sectors are exchanged by charge conjugation, and the in-use instruments are assumed to preserve the reference charge sector.  The proof is elementary; its role is to make the operational assumptions explicit: it identifies exactly which statements survive tracing out the reference, which require a charge-conjugation-invariant data input, and which fail if the reference is retained or if a reset is charge-conjugation breaking.

Our second contribution is a model-conditional error statement.  We represent label, reference, walking, and selected coherence failures by a common classical flag refinement of an accepted run.  The probability of any bad flag is bounded by a union bound.  This gives a scalar target-infidelity budget, not a device-level logical error rate. We separately prove a channel-distance stability bound under explicit complete-instrument assumptions; the scalar budget parameters are not automatically bounds in that metric.  The distinction matters because the registered coherence term $\gamma/2$ refers to a specified measurement test; for another route, such as the two-qutrit auxiliary-state route, the correct fidelity term is different.

The paper has four deliberate boundaries.  First, ``universal'' always means dense single-qutrit closure in the registered ideal sector model and on the stated measurement records.  Second, the ternary Golay calculation is separated from the abstract syndrome/recovery decoder: the former has the F2 postselection factor $1/1728$, while the latter accepts every abstract syndrome and has no native charge-$4e$ extraction circuit.  Third, all F4 and F5-S feasible regions are abstract parameter or synthetic-data results.  Fourth, the literature already contains charge-$4e$ proposals, ternary Golay distillation, and measurement-assisted anyonic universality; the contribution here is their audited reference-assisted closure and error-interface package.  No material, device, or experimental data are used.

The main results are organized as follows.  Section~\ref{sec:model} fixes the qutrit and reference conventions.  Section~\ref{sec:closure} proves the closure theorem.  Section~\ref{sec:algebra} summarizes the registered braid and reference algebra, and Section~\ref{sec:universality} reproduces the known ideal single-qutrit density result in our conventions.  Section~\ref{sec:magic} describes state preparation, injection, the ternary Golay calculation, and the abstract $[[11,1,5]]_3$ decoder.  Section~\ref{sec:error} gives the conditional budget theorem.  Section~\ref{sec:synthetic} records the synthetic-data interface and its limits.  We conclude with the experimental and mathematical problems that remain open.

\section{Charge-conjugation qutrit and operation model}
\label{sec:model}

Let $\omega=e^{2\pi i/3}$ and let $\D\cong\mathbb C^3$ be the logical data qutrit with computational basis $\ket{j}$, $j\in\mathbb Z_3$.  The Weyl operators are denoted by $X$ and $Z$, with
\begin{equation}
 X\ket{j}=\ket{j+1},\qquad Z\ket{j}=\omega^j\ket{j},\qquad ZX=\omega XZ.
\end{equation}
Charge conjugation is the unitary
\begin{equation}
 \C\ket{j}=\ket{-j},\qquad \C X\C=X^{-1},\qquad \C Z\C=Z^{-1}.
\label{eq:C}
\end{equation}
The same notation is used for the corresponding operation on a reference qutrit $\R$. Write $Q=\operatorname{diag}(1,\omega^2,\omega^2)$ and $H_{ab}=\omega^{ab}/\sqrt3$ for the registered braid generators, so $H^2=\C$. We use the sector operator $Z_R$ and restrict the usable reference to
\begin{equation}
 S_R=\operatorname{span}\{\ket{1}_R,\ket{2}_R\}.
\end{equation}
The two coherent pure states used below are
\begin{equation}
 \ket{+}_R=\frac{\ket{1}_R+\ket{2}_R}{\sqrt 2},\qquad
 \ket{-}_R=\frac{\ket{1}_R-\ket{2}_R}{\sqrt 2}.
\end{equation}

The reference basis labels the Abelian charge of the first vortex pair in a
neutral four-defect register: $k=0$ is the vacuum channel, while $k=1,2$
are exchanged by charge conjugation. The registered ideal preparation
starts from $\ket{0}_R$, applies $H_R$, and accepts the nonzero outcome of
the binary test $T(Z_R)$. This outcome has probability $2/3$ and prepares
$\ket{+}_R$ if the test preserves the two nonzero sectors' coherence.
If a hidden environment instead records which nonzero charge occurred,
the prepared reference is $(\ket{1}\bra{1}+\ket{2}\bra{2})/2$.
Both preparations have balanced populations and satisfy the reduced-data
closure hypotheses; only the first supplies coherence to a route that
explicitly requires it.

Alignment is a relational condition distinct from balance. All data may
share one reference. For two references, the registered measurement of
$Z_RZ_{R'}^{-1}=1$ separates the same-sector span
$\operatorname{span}\{\ket{11},\ket{22}\}$ from the opposite-sector span
$\operatorname{span}\{\ket{12},\ket{21}\}$; applying
$C_{R'}=H_{R'}^2$ to the latter aligns the two references without revealing
the absolute charge. This preparation correction precedes the restriction
to operations preserving the in-use reference sectors. These are ideal
primitive-level constructions, not measured device preparation fidelities.

An elementary visible instrument has a classical result $a$ and a completely positive (CP), trace-nonincreasing map $\I_a$.  In the reference basis it is required to be sector preserving.  Thus, on $S_R$, its Kraus operators have the form
\begin{equation}
 K_{a,\mu}=K_{a,\mu,1}\otimes\ket{1}\bra{1}_R+K_{a,\mu,2}\otimes\ket{2}\bra{2}_R,
\label{eq:block}
\end{equation}
where $\mu$ is a hidden Kraus label and $\I_{a,k}(\rho)=\sum_\mu K_{a,\mu,k}\rho K_{a,\mu,k}^\dagger$. Covariance is imposed on these CP maps, without requiring a particular matching of Kraus decompositions. The central assumption is the same-label relation
\begin{equation}
 \I_{a,2}=\Ad_{\C}\circ\I_{a,1}\circ\Ad_{\C^{-1}}.
\label{eq:samelabel}
\end{equation}
Adaptive choices may depend on earlier visible records, but not on the unobserved reference sector.  The phrase ``same label'' is essential.  If charge conjugation permutes visible labels, same-label covariance requires summing CP maps over complete label-permutation orbits and making feedback depend on the resulting quotient record; a mere relabeling need not suffice.

\section{Reference-assisted closure theorem}
\label{sec:closure}

We now state the central result.  For a visible record $r=(a_1,\ldots,a_t)$, write $\I_{r,k}$ for the composed CP map in reference sector $k\in\{1,2\}$.  Let $\sigma_R$ be a reference state supported on $S_R$, with populations $w_k=\bra{k}\sigma_R\ket{k}$ and coherence $z=\bra{1}\sigma_R\ket{2}$.

\begin{theorem}[Reference-assisted closure and record blindness]
\label{thm:closure}
Assume sector preservation~\eqref{eq:block}, same-label covariance~\eqref{eq:samelabel}, and feedback that depends only on visible records.  Then:
\begin{enumerate}
 \item For every adaptive record $r$,
 \begin{equation}
  \I_{r,2}=\Ad_{\C}\circ\I_{r,1}\circ\Ad_{\C^{-1}}.
 \label{eq:recordcov}
 \end{equation}
 \item If the data input obeys $\C\rho\C^{-1}=\rho$, every record probability is independent of $\sigma_R$ inside $S_R$ and equals
 \begin{equation}
  p(r)=\Tr\I_{r,1}(\rho)=\Tr\I_{r,2}(\rho).
 \end{equation}
 For $p(r)>0$, after tracing out the reference, the normalized data state is independent of $z$ but generally depends on $w_1,w_2$:
 \begin{equation}
  \rho_{D|r}=\frac{w_1\I_{r,1}(\rho)+w_2\I_{r,2}(\rho)}{p(r)}.
 \label{eq:conditionalstate}
 \end{equation}
 \item For arbitrary $\rho$, the unnormalized reduced branch is
 \begin{equation}
  \Tr_R\I_r(\rho\otimes\sigma_R)=w_1\I_{r,1}(\rho)+w_2\I_{r,2}(\rho).
 \label{eq:mixture}
 \end{equation}
 If the sector maps are unitary channels (or share a scalar branch probability, removed on conditioning), this is the channel mixture
 \begin{equation}
  w_1U_r\rho U_r^\dagger+w_2(\C U_r\C^{-1})\rho(\C U_r\C^{-1})^\dagger.
 \end{equation}
 Equal weights give the symmetry average of the channel $\Ad_{U_r}$ under conjugation by $\C$. This is not in general the state twirl $\rho\mapsto(\rho+\C\rho\C^{-1})/2$. Equal weights must be prepared or imposed by a specified symmetrization; they do not follow from tracing out $R$.
 \end{enumerate}
\end{theorem}

\begin{proof}
For one primitive, sector preservation gives a block decomposition of every Kraus operator.  Equation~\eqref{eq:samelabel} relates the diagonal CP maps. Composition preserves that relation because adjacent $\Ad_{\C^{-1}}\circ\Ad_{\C}$ factors cancel; no Kraus-by-Kraus covariance is required.  For an adaptive protocol, induction on the visible record tree applies because the next operation is selected from the already observed record and is therefore the same operation in both sectors.  This proves~\eqref{eq:recordcov}.

Expanding the joint state in reference matrix units, the diagonal blocks after record $r$ are $w_k\I_{r,k}(\rho)$ and the off-diagonal blocks are proportional to $z$ and $z^*$.  The partial trace over $R$ removes the off-diagonal blocks.  If $\rho$ is $\C$-invariant, equation~\eqref{eq:recordcov} makes the two diagonal traces equal, which proves record blindness and~\eqref{eq:conditionalstate}.  No $\C$-invariance is needed for~\eqref{eq:mixture}.
\end{proof}

\begin{remark}[What the theorem does not say]
The theorem does not assert that a coherent reference becomes a classical mixture when it is retained.  The identity protocol with $\ket{+}_R$ gives $\rho\otimes\ket{+}\bra{+}_R$, whose off-diagonal reference block is nonzero.  An inaccessible environment can record the sector while preserving each reference sector and both visible diagonal CP maps. It can therefore dephase a retained reference without breaking visible record blindness. Reading that environmental label, or applying a charge-conjugation-breaking reset, can instead violate same-label covariance. Sector transitions, hidden dephasing, and visible charge readout are distinct mechanisms.  Finally, if $\C$ sends a visible result $a$ to $\pi(a)$, for a $\C$-invariant data input the direct conclusion is the paired-record identity $p(a|1)=p(\pi(a)|2)$; same-label record blindness requires an invariant record quotient.
\end{remark}

\subsection{Finite-depth stability and zero-sector contamination}
\begin{theorem}[Stability for complete labelled instruments]
\label{thm:stability}
Include the visible history, feedback controller, and abort status in the
output memory, and pad halted branches by identity steps up to depth $T$.
Let $\Phi_t,\widetilde\Phi_t$ be the ideal and actual complete
trace-preserving instrument channels, including all outcomes. Assume a
uniform bound over histories
\[
 \tfrac12\|\widetilde\Phi_t-\Phi_t\|_\diamond\le\epsilon_t,
 \qquad D(\widetilde\rho_0,\rho_0)\le d_0,
 \qquad D(\rho,\sigma)=\tfrac12\|\rho-\sigma\|_1.
\]
The ideal instruments satisfy Theorem~\ref{thm:closure}; where leakage
spaces are retained, the comparison uses an explicit trace-preserving
extension of the ideal maps to those spaces. Then complete outputs and
record laws differ by at most
\[
 \zeta=\min\{1,d_0+\textstyle\sum_{t=1}^T\epsilon_t\}.
\]
For two supported reference preparations and a common
$C$-invariant data input, actual record laws therefore have total variation
at most $\min\{1,\zeta_\sigma+\zeta_\tau\}$.
For an acceptance event with actual and ideal probabilities $p,q>0$, the
normalized accepted states satisfy
\[
 D(\widetilde\rho_A,\rho_A)\le
 \min\{1,\zeta/\max(p,q)\}.
\]
\end{theorem}
\begin{proof}
Trace-distance contractivity and the channel bound give the telescoping
estimate $D_t\le D_{t-1}+\epsilon_t$; classical readout is also contractive.
Compare the two actual record laws through their common ideal law.
For accepted subnormalized states $x,y$ of traces $p,q$, the rejected flag
block implies $\|x-y\|_1+|p-q|\le2\zeta$. Expanding $x/p-y/q$ first with
denominator $p$, then with denominator $q$, gives the stated bound.
\end{proof}

This is a channel-distance assumption, not a consequence of a small
commutator with $Z_R$. For example, visible sector measurements have
Kraus operators $I_D\otimes\ket{k}\bra{k}$ that commute exactly with
$Z_R$, yet distinguish $k=1$ from $k=2$ perfectly and violate same-label
covariance. No device value of $\epsilon_t$ is established here. For an
unbounded stopping protocol one needs a stopping-time bound or an
explicit truncation and overflow contribution; its mean length alone
does not justify replacing $T$ in the estimate.

If a prepared reference has zero-sector weight
$q_0=\bra0\widetilde\sigma_R\ket0<1$, its distance from its normalized
projection onto $S_R$ is at most $\sqrt{q_0}$, providing a preparation
contribution to $d_0$. For block-incoherent zero-sector contamination this
contribution is exactly $q_0$. If all three reference sectors are
preserved, the reduced branch acquires $w_0\mathcal I_{r,0}(\rho)$, whose
record probabilities need not equal those of the nonzero sectors even
when the zero block is charge-conjugation covariant. The original
record-blindness theorem thus remains restricted to $S_R$.

\section{Registered algebra and the source of the Pauli sectors}
\label{sec:algebra}

The charge-$4e$ defect data used here are the $\mathbb Z_3$ charge-conjugation defect data of the parafermionic-vortex proposal \cite{Shi2026}.  The defect has quantum dimension $\sqrt 3$ and the defect fusion channel contains the three Abelian charge sectors.  The braid matrices and probe words were reconstructed from the registered $F$ and $R$ data, including the path convention for the five braid diagrams.  The resulting four- and eight-strand checks reproduce the source matrices and the charge-conjugation mirror relations.  The use of $G$-crossed language follows the standard defect formalism \cite{Barkeshli2015}. More general parafermion encodings and available braid sets can realize different Clifford images~\cite{Hutter2015}; the following order-24 statement concerns our specified four-defect generators.

For the fixed braid generators used in the source construction, the single-qutrit phase image generated by the registered $Q$ and $H$ has order
\begin{equation}
 |\langle Q,H\rangle/\text{phases}|=24.
\end{equation}
This is the single-qutrit Clifford image without the nontrivial Pauli translations.  The braid-only representation therefore should not be described as the full Clifford group or as generating every stabilizer state without an additional sector argument.

The reference qutrit supplies that sector argument.  In the $k=1$ and $k=2$ reference blocks, the registered operations are charge-conjugate copies.  Each block has the same Clifford image, while the reference-assisted controlled operations and measurement projectors supply the nontrivial Pauli translations.  The generators normalize the qutrit Weyl group. Their enumeration agrees with the projective Clifford structure $\mathbb F_3^2\rtimes SL(2,3)$, of order $9\cdot24=216$~\cite{Appleby2005}, in each of these two sectors; the $k=0$ block is a smaller linear subgroup.  This is a conditional closure statement: it assumes the reference is available, remains in the registered sector model, and is used with the specified visible records.

The distinction between the 24-element braid image and the 216-element reference-assisted image is operationally important.  An external data state does not automatically acquire an equal twirl.  By Theorem~\ref{thm:closure}, it sees the population-weighted mixture in equation~\eqref{eq:mixture}.  Equal weights require a prepared balanced reference or an explicit symmetrization.

\section{Self-contained density check for the registered gate set}
\label{sec:universality}

The density result for this gate family is known: Cui and Wang~\cite{CuiWang2015}, Lemma~4, give the same commutator and eigenvalues used below. We supply a self-contained check tied to the registered matrices, without claiming a new density theorem.  Let $C_1=\langle Q,H,X,Z\rangle$ denote the 216-element projective qutrit Clifford group in the reference-supported sector, and let
\begin{equation}
 F_0=\operatorname{diag}(-1,1,1).
\end{equation}
The three measured N3 branches are Pauli conjugates of $F_0$. The outcome identifies the branch, but two-sided conjugacy does not furnish an output-only Pauli correction on an arbitrary input: for a branch $F_j$, such a correction would have to be $F_0F_j$, which is non-Weyl for $j\ne0$. Realizing a designated Flip uses the registered repeat-until-target protocol and retains its full branch record.  Direct conjugation gives
\begin{equation}
 F_0XF_0=\begin{pmatrix}0&0&-1\\-1&0&0\\0&1&0\end{pmatrix}.
\end{equation}
Its normalized shift-column phase ratios are $(1,-1,1)$, which are not the Weyl phases $(1,\omega^b,\omega^{2b})$.  Thus $F_0$ is genuinely non-Clifford.

Define $D=\operatorname{diag}(1,1,\omega^2)$ and $P_2=DF_2$, with $F_2=X^{-1}F_0X$.  Up to a scalar, $D=Q(XQX^{-1})$, so $P_2$ belongs to $\langle C_1,F_0\rangle$.  For
\begin{equation}
 W=HP_2H^\dagger P_2^\dagger
\end{equation}
we find exactly
\begin{equation}
 \chi_W(\lambda)=(\lambda-1)(\lambda^2-\tfrac43\lambda+1),\qquad \cos\theta=\tfrac23.
\end{equation}
The eigenvalue $e^{i\theta}$ is not a root of unity, since $e^{i\theta}+e^{-i\theta}=4/3$ would then be an algebraic integer.  Therefore the closure contains the one-parameter subgroup generated by $A_0=(W-W^\dagger)/(2i\sin\theta)$.  In the Weyl basis $W_{ab}=\omega^{2ab}X^aZ^b$, define $c_{ab}=\Tr(W_{ab}^\dagger A_0)/3$. The eight coefficients are
\begin{align}
 c_{01}&=\frac{\sqrt{15}-3i\sqrt5}{20},& c_{02}&=\frac{\sqrt{15}+3i\sqrt5}{20},\nonumber\\
 c_{10}&=-\frac{\sqrt{15}+3i\sqrt5}{20},& c_{20}&=-\frac{\sqrt{15}-3i\sqrt5}{20},\nonumber\\
 c_{11}&=\frac{\sqrt{15}}{30},&c_{22}&=\frac{\sqrt{15}}{30},\nonumber\\
 c_{12}&=\frac{\sqrt{15}-3i\sqrt5}{60},&c_{21}&=\frac{\sqrt{15}+3i\sqrt5}{60}.
\end{align}
Every coefficient is nonzero. Since
\begin{equation}
 W_{uv}W_{ab}W_{uv}^\dagger=\omega^{va-ub}W_{ab},
\end{equation}
the finite Fourier character matrix shows that the nine Hermitian conjugates of $A_0$ have complex span of dimension eight. Complex and real ranks agree for Hermitian generators, so their real span is all traceless Hermitian matrices. Let $K$ be the closed generated subgroup for determinant-one representatives. The irrational rotation above puts $\exp(itA_0)$ in $K$, and hence $iW_{uv}A_0W_{uv}^\dagger$ in its Lie algebra. This Lie algebra is therefore $\mathfrak{su}(3)$. The identity component of the closed Lie subgroup $K\subseteq SU(3)$ is all of the connected group $SU(3)$; thus
\begin{equation}
 \overline{\langle C_1,F_0\rangle}=SU(3)
\end{equation}
with that representative convention, and the projective single-qutrit gate set is dense.

This is a sector-wise ideal statement.  It assumes the reference sector is aligned and visible records and the target-Flip protocol are retained. Multi-qutrit universality would additionally require a usable entangling gate, such as SUM, and consistent reference and record semantics for the composite circuit~\cite{CuiWang2015}.  It does not claim that a reduced channel with an unobserved reference sector is unitary, nor does it provide a fault-tolerant compiler.  As a finite reproducibility check, the word $W^{3005}$ approximates $\exp(i\pi A_0/2)$ with circular angle error $6.5244\times10^{-5}$ and Frobenius error below $9.23\times10^{-5}$; the unreduced accounting is 48,080 Clifford symbols and 6,010 Flip symbols.

\section{Non-stabilizer states, injection, and ternary Golay distillation}
\label{sec:magic}

Define
\begin{equation}
 \ket{\pm}=\frac{\ket{1}\pm\ket{2}}{\sqrt2}.
\end{equation}
In the phase-point convention $A_{q,p}=\sum_{x\in\mathbb F_3}\omega^{px}\ket{q+x/2}\bra{q-x/2}$, $W_\rho(q,p)=\Tr(A_{q,p}\rho)/3$, the two states have minima~\cite{Gross2006}
\begin{equation}
 \min W_{-}=-\frac13,\qquad \min W_{+}=-\frac16,
\end{equation}
so both are non-stabilizer states.  The exact Wigner tables and the preparation identities are part of the F1/F2 calculation.  For example, the registered SUM/controlled operations followed by the two-dimensional $M_2$ measurement produce $\ket{+}$ and $\ket{-}$ branches with equal conditional weights in the reference-supported sector.  The interferometric and coherent-test routes then supply the state needed by the injection circuits, with the outcome-dependent charge-conjugation corrections implemented by the reference walk.

The qutrit walk has three equally likely flip labels in the ideal model.  If the record is read without error, the expected number of rounds to reach a designated target label is $3$.  With a total readout error rate $m$, two symmetric wrong labels, and the registered target-stop rule, the exact absorption calculation gives
\begin{equation}
 r_m=\frac{m(6-m)}{2+5m-m^2}.
\label{eq:walk}
\end{equation}
At small $m$, this behaves as $3m+O(m^2)$.

For distillation, the registered $\G_1=\langle Q,H\rangle$ twirl gives the strange-state family
\begin{equation}
 \rho_\delta=(1-\delta)\ket{-}\bra{-}+\delta I/3,\qquad \epsilon=1-\bra{-}\rho_\delta\ket{-}=2\delta/3.
\end{equation}
This full 24-element twirl preserves strange-state fidelity and mixes its rank-two orthogonal complement. The two-element charge-conjugation twirl alone does not produce this family: it leaves $\ket0\bra0$ unchanged. We reproduce the ternary-Golay strange-state distillation map of Prakash~\cite{Prakash2020} in these parameter conventions, including its cubic suppression, crossing, and ideal acceptance.  For eleven independent, identically prepared copies $\rho_\delta^{\otimes11}$, the one-round output parameter has the form
\begin{equation}
 \delta_{\rm out}(\delta)=\frac{\delta^3P(\delta)}{2Q(\delta)},
\end{equation}
where $P$ and $Q$ are the exact ternary-Golay enumerator polynomials~\cite{Prakash2020}. Independent integer-enumerator reconstruction and exact polynomial root counting find one nonzero crossing in $0<\delta<3/4$, in the bracket:
\begin{equation}
 \delta_*\in(0.38714,0.38716),\qquad
 \epsilon_*:=\frac{2\delta_*}{3}=0.2581028976478637.
\label{eq:threshold}
\end{equation}
The corresponding accepted-syndrome probability has constant term $1/1728$ at the ideal point, and the leading output suppression is cubic.  For the untwirled rank-two family $\rho_\epsilon=(1-\epsilon)\ket{-}\bra{-}+\epsilon\ket{+}\bra{+}$, the one-round fidelity map has a unique crossing in the registered range $0<\epsilon<1/2$:
\begin{equation}
 \epsilon_u^*=\frac{3-\sqrt5}{4}=0.1909830056251.
\end{equation}
The other physical fixed points of the untwirled map are listed in Appendix~\ref{app:golay}. Independent input copies and the full $\G_1$ twirl are parts of the stated input model for equation~\eqref{eq:threshold}; local twirling alone does not remove arbitrary correlations between copies.

The gate-density conclusion here concerns the explicit Clifford-plus-Flip set of Section~\ref{sec:universality}; we do not assume that every non-stabilizer state is distillable or that every non-Clifford extension has the same operational realization. Broader prime-dimensional distillation constructions provide context~\cite{Campbell2012}. The present calculation reproduces a specified input family and one-round map. It does not construct a native 11-qutrit Golay extraction circuit or show that arbitrary device noise automatically has this form.

\subsection{Abstract ternary-Golay syndrome and recovery decoder}
The classical ternary Golay code generated by $g(x)=x^5+2x^3+x^2+2x+2$ gives an exact CSS $[[11,1,5]]_3$ code.  An independent stabilizer check matrix is
\begin{equation}
H=\begin{pmatrix}
1&0&1&2&2&2&1&0&0&0&0\\
1&1&1&0&1&1&0&1&0&0&0\\
2&1&0&2&1&2&0&0&1&0&0\\
1&2&2&2&1&0&0&0&0&1&0\\
0&1&2&2&2&1&0&0&0&0&1
\end{pmatrix}.
\end{equation}
For a Pauli label $E(a,b)=X^aZ^b$, use the syndrome labels $(Ha,Hb)$ over $\mathbb F_3$, with the measured $X$-check eigenvalue sign absorbed into the second label.  For each CSS component separately, the decoder stores, for each of the 243 classical syndromes, the lexicographically first minimum-weight representative $r(s)$ and applies $X^{-r(Ha)}Z^{-r(Hb)}$.  The 243 classical ternary errors of weight at most two have distinct syndromes, the syndrome representative histogram is $(1,22,220)$ by weight, and every syndrome coset has size $3^6$.  The joint quantum decoder has $243^2=3^{10}$ syndrome pairs. Every Pauli error of support weight at most two has each component of classical weight at most two, so all $1+88+3520=3609$ such Pauli labels are corrected to the identity logical sector.

Let $p$ be the total nonidentity probability at one site in the independent Pauli model: the identity has probability $1-p$, and each of the eight other labels has probability $p/8$. This $p$ is distinct from the strange-state depolarizing parameter $\delta$. If $A_{uv,w}$ counts support-weight-$w$ errors with residual logical label $(u,v)$, then
\begin{equation}
 P_{uv}(p)=\sum_{w=0}^{11}A_{uv,w}(1-p)^{11-w}(p/8)^w.
\end{equation}
The exact dynamic program over $9^{11}=31{,}381{,}059{,}609$ labels obeys $\sum_{uv}A_{uv,w}=\binom{11}{w}8^w$. The identity row begins $(1,88,3520,23760,\ldots)$, giving the guaranteed-correctable mass
\begin{equation}
 P_{\le2}(p)=(1-p)^{11}+11p(1-p)^{10}+55p^2(1-p)^9\le P_{00}(p).
\label{eq:decoderbound}
\end{equation}
All nonidentity logical rows vanish through weight two. Directly enumerating the 84,480 weight-three labels gives 23,760 identity-logical recoveries and 60,720 failures, so the bound is not tight. For example, $Z_1Z_2Z_3$ receives the recovery $Z_7Z_{10}$, leaving a nontrivial weight-five logical residual. The classical covering radius is two; the joint Pauli-syndrome covering radius is four, with minimum-weight histogram $(1,88,3520,47520,7920)$. Neither covering radius is the guaranteed correction radius.

The ideal full-syndrome recovery instrument rejects no syndrome, so its acceptance is one. In contrast, strange-state distillation projects a particular eleven-copy input onto the zero-syndrome code space. For $\ket{-}^{\otimes11}$, the stabilizer weight enumerator is
\begin{equation}
 B(t)=1+528t^6+7920t^8+11000t^9+23760t^{10}+15840t^{11}.
\end{equation}
Using the phase-symmetrized $W_{ab}=\omega^{2ab}X^aZ^b$ convention, each nonidentity single-site Weyl expectation is $-1/2$, hence
\begin{equation}
 A(0)=\frac{B(-1/2)}{3^{10}}=\frac{2187/64}{3^{10}}=\frac1{1728}.
\end{equation}
An independent CSS overlap calculation gives the signed numerators $(0,12,-12)$ and the same value $(12^2+12^2)/(3^5 2^{11})$. These two acceptances refer to different instruments and inputs; they cannot be identified or multiplied without a specified composed protocol. All-syndrome acceptance does not imply costless or faultless physical measurement. A native charge-$4e$ extraction schedule, leakage model, and correlated-noise threshold remain open.

\section{Error channels and an accepted-run budget}
\label{sec:error}

The F3/F4 model separates the following quantities: visibility and static phase parameters $(\sigma_\phi,\Delta)$, a silent $\pi$-jump probability $\eta$, reference-flip parameter $\nu$, walk readout error $m$, coherence parameter $\gamma$, and a herald success parameter $q$.  In the iid-probe model, the label error for an odd number $N$ of probes is computed from the same visibility-and-probe logic used in standard anyon interferometry \cite{Bonderson2008}:
\begin{align}
 V&=e^{-\sigma_\phi^2/2},\\
 p_{\rm pr}&=(1+V\cos\Delta)/2,\\
 p_\eta&=(1-\eta)p_{\rm pr}+\eta(1-p_{\rm pr}),
\end{align}
followed by the majority tail of a binomial distribution.  In the persistent single-jump model, the silent-jump analysis gives the registered critical value
\begin{equation}
 \eta_c=0.0778691008147,
\end{equation}
and the optimal odd probe number changes across this boundary.  The iid-probe model and the robust maximum of the iid and persistent models are treated separately below; the value $\eta_c$ is not a universal threshold for arbitrary drift or Markov noise.

Reference redundancy uses the conditional no-loss flip rate
\begin{equation}
 f=\frac{\nu}{2-\nu},\qquad r_\nu=3f^2-2f^3.
\label{eq:referenceerror}
\end{equation}
The loss probability is not included in $r_\nu$; it must be included separately if an abort is counted as a failed attempt rather than conditioned away.

The scalar budget requires a precise event interpretation.  Let $A$ be the accepted-run event and let $L,R,W,D$ denote label, reference, walk, and selected coherence bad flags.  For the normalized accepted output $\rho_{\rm out}$, we use squared fidelity to a pure target, $F_{\rm out}=\bra{\psi_{\rm ideal}}\rho_{\rm out}\ket{\psi_{\rm ideal}}$. We assume that the accepted protocol has a common classical flag refinement, that each stated term bounds its bad flag conditional on the same acceptance event, and that the all-good flag produces this pure target. These conditions must be checked for each application; a marginal diagnostic fidelity alone does not supply such a flag model.  We then obtain the following proposition.

\begin{proposition}[Model-conditional target-infidelity budget]
Under the accepted-run flag assumptions above, with $e_N=\sup_i\Pr(L\mid A,i)$, equations~\eqref{eq:walk} and~\eqref{eq:referenceerror}, and a coherence term $r_\gamma$ defined for the chosen route,
\begin{equation}
 1-F_{\rm out}\leq \Pr(L\cup R\cup W\cup D\mid A)
 \leq e_N+r_\nu+r_m+r_\gamma.
 \label{eq:budget}
\end{equation}
No independence among the four bad events is required for the final union bound.
\end{proposition}

\begin{proof}
On the all-good flag the output has unit fidelity with the ideal output.  Every bad flag contributes at most one unit of infidelity, so the total infidelity is bounded by the probability of the union.  The union bound gives the second inequality.  Independence is used only when deriving the closed forms for individual terms.
\end{proof}

The coherence term is route dependent.  For the specified H6(a) $\ket{+}$ measurement test, $r_\gamma=\gamma/2$.  For the N3 CP-map route, the registered map is independent of $\gamma$, so $r_\gamma=0$.  For the N2 auxiliary-state route, the exact fidelity is
\begin{equation}
 F_{N2}(\gamma)=\frac{3-\gamma}{3(1+\gamma)},\qquad
 r_\gamma^{\rm N2}=1-F_{N2}=\frac{4\gamma}{3(1+\gamma)}.
\label{eq:n2gamma}
\end{equation}
Consequently $\gamma/2$ must not be called a universal trace or diamond distance.  The channel stability theorem (Theorem~\ref{thm:stability}) instead assumes complete-instrument diamond-distance bounds. No conversion from the scalar $r_\gamma$ values or the F4 grid to those bounds is established here.

The crossing in equation~\eqref{eq:threshold} applies to eleven independent inputs in the $\G_1$-twirled family, including the designated diagonal family only after that full twirl. Without twirling, the binary diagonal family instead has the smaller crossing $\epsilon_u^*$.  It does not automatically cover retained reference coherence, leakage, correlated Markov jumps, continuous drift, thermal quasiparticles, or decoder faults.

\section{Abstract budget regions and synthetic-data protocol}
\label{sec:synthetic}

The F4 interface evaluated the registered scalar budget on 28,800 abstract parameter combinations.  Under the original H6(a)-style coherence convention, 16,528 points satisfy $\epsilon_{\rm end}<\epsilon_*$, or $57.389\%$.  A later reconciliation of the independent-iid and persistent single-jump models gives
\begin{center}
\begin{tabular}{lcc}
\toprule
label process & feasible points (abstract model) & fraction\\
\midrule
iid probe & 16,528 & 57.389\%\\
single jump & 10,524 & 36.542\%\\
robust maximum & 10,524 & 36.542\%\\
\bottomrule
\end{tabular}
\end{center}
The robust row covers only the two registered jump models.  It is not a general noise theorem.

F5-S specifies how a future experiment could fit the interface without silently treating correlated data as iid.  The synthetic planning run contains 754,000 unique ordinary rows, 854,000 logical-channel rows, and 2,500,000 ordered jump-probe events, for about 3.254 million deduplicated raw events.  It uses exact or high-precision formula checks, planning 99\% interval fields (with no empirical coverage claim), held-out iid and single-jump semantic diagnostics, and a Markov negative control that must remain inconclusive.  The model-selection smoke tests classified all six iid examples as iid, all six persistent-jump examples as single-jump, and all six Markov examples as inconclusive.

These numbers describe a statistical protocol, not an experiment.  A real run must retain event ordering, readout scores, timeouts, loss flags, postselection outcomes, and enough independent blocks to estimate effective sample size.  The parameter $q$ is a resource/postselection interface; it is not a measured device success probability in the present paper.

\section{Scope, experimental interface, and open problems}
\label{sec:limits}

The theory establishes a conditional chain:
\begin{align*}
 &\text{registered charge-conjugation operations}\\
 &\quad\Longrightarrow\text{reference-assisted Clifford sectors}\\
 &\quad\Longrightarrow\text{non-stabilizer injection}\\
 &\quad\Longrightarrow\text{model-conditional one-round distillation}.
\end{align*}
Each arrow carries assumptions.  The closure theorem applies to sector-preserving instruments and same-label covariance.  The gate algebra assumes the ideal registered operation set.  The resource calculation assumes the specified measurement records and postselection.  The budget theorem assumes a common classical failure flag and accepted-run conditioning.

The next experimental step is therefore a parameter-mapping exercise, not a claim that the F4 grid has already been realized.  A device characterization should estimate visibility and phase drift from the three-point interferometric calibration; estimate $\eta$ from time-ordered jump probes; estimate $\nu$ with explicit loss accounting; estimate $m$ from the target-stop walk; and fit $\gamma$ separately for the H6(a), N2, and N3 routes.  A held-out model-selection block should reject the iid/single-jump abstraction when a Markov or drift model is favored.  The resulting joint uncertainty should be propagated through equation~\eqref{eq:budget}, rather than by multiplying marginal 99\% intervals.

Several theory problems remain open.  A native 11-qutrit ternary-Golay decoding circuit should be constructed and its error propagation analyzed.  The complete-instrument stability theorem needs device-specific bounds, an explicit leakage space, and a stopping-time or truncation analysis for unbounded protocols.  The route-dependent budget should be recast in a common channel metric if a fault-tolerance theorem is desired.  Finally, the charge-conjugation reference channel should not be identified with the dynamically gauged $\mathbb Z_3\rtimes\mathbb Z_2$ model studied in the separate G campaign: the present charge-conjugation defect has dimension $\sqrt3$, whereas the transposition-flux anyon in the gauged $D(S_3)$ theory has dimension $3$. Quantum-double circuit work~\cite{Chen2025,CuiHongWang2014} is therefore a comparison, not a direct hardware implementation of this defect model.

\section{Conclusion}

We have written the charge-$4e$ qutrit route as a conditional, auditable theory rather than as an unconditional universal-fault-tolerance claim.  The central structural result is the reference-assisted closure theorem, which separates record blindness, reduced-channel mixing, and retained-reference coherence.  The registered operation algebra and reproduced ternary Golay calculation then provide a conditional route from charge-conjugation defects to non-stabilizer qutrit resources.  The accepted-run budget quantifies which error assumptions are being made and exposes the route dependence of the coherence term.  The remaining work is a native decoder construction and an experimental parameter determination.

\section*{Acknowledgments}

The research program was designed and directed by the author. Derivations, computations, code, figures and drafting used AI assistance (Anthropic Claude), and computational checks and AI-assisted review used OpenAI Codex. These checks do not constitute independent human peer review. The author takes full responsibility for the content. This work received no external funding.

\appendix
\section{Golay polynomials and exact conventions}
\label{app:golay}

For reproducibility, the polynomials used in equation~\eqref{eq:threshold} are listed here in ascending powers of $\delta$:
\begin{align}
 P(\delta)={}&13365-71280\delta+181764\delta^2-283536\delta^3
 +292710\delta^4\\
 &-203280\delta^5+92180\delta^6-24816\delta^7+3021\delta^8,\\
 Q(\delta)={}&2187-16038\delta+53460\delta^2-102465\delta^3
 +121770\delta^4\\
 &-86328\delta^5+23628\delta^6+18810\delta^7-25245\delta^8\\
 &+13750\delta^9-3960\delta^{10}+495\delta^{11}.
\end{align}
The ideal acceptance expansion is
\begin{equation}
 A(\delta)=\frac{1}{1728}-\frac{11}{2592}\delta+\frac{55}{3888}\delta^2+O(\delta^3).
\end{equation}
The acceptance is exactly $A(\delta)=Q(\delta)/(64\cdot3^{10})$, and $\delta_{\rm out}=55\delta^3/18+O(\delta^4)$. The v0.3 independent validation reconstructs both integer polynomials and uses exact root counts, independently reproducing the original exact twirled calculation. For the untwirled binary family it separately enumerates all $2^{11}$ signed CSS input patterns, strengthening the historical floating-point/rationalized check. Writing $N_u$ for the unnormalized target overlap and $A_u$ for acceptance, the crossing polynomial factors as
\begin{align}
 N_u(e)-(1-e)A_u(e)={}&-\frac{e(e-1)(2e-1)}{5184}
 (4e^2-6e+1)(4e^2-2e-1)\nonumber\\
 &\times(8e^4-16e^3+20e^2-12e+3).
\end{align}
Here $e=\epsilon$. The binary input family is physical for $0\le e\le1$, whereas the registered comparison uses $0\le e\le1/2$. The acceptance $A_u(e)$ is strictly positive on $[0,1]$: its nonnegative Bernstein coefficients give $A_u(e)\ge[(1-e)^{11}+e^{11}]/1728>0$. The quartic factor is
\[
 8(e-\tfrac12)^4+8(e-\tfrac12)^2+\tfrac12>0,
\]
so the complete set of real roots is
\[
 \left\{\frac{1-\sqrt5}{4},\;0,\;\frac{3-\sqrt5}{4},\;
 \frac12,\;\frac{1+\sqrt5}{4},\;1,\;\frac{3+\sqrt5}{4}\right\}.
\]
Only $(3-\sqrt5)/4$ lies in the open registered half-range. The other interior physical fixed points are $e=1/2$ and $(1+\sqrt5)/4$. At $e=1/2$, the binary state is $\operatorname{diag}(0,1/2,1/2)$, not $I/3$. For the fixed target $\ket{-}$, the sign of $F_u(e)-(1-e)$, where $F_u=N_u/A_u$, is successively $+,-,+,-$ on the four intervals separated by these three interior roots. Thus the root $(1+\sqrt5)/4$ also changes the direction of target-infidelity improvement; it is outside the registered half-range. These checks certify the finite-code, product-input maps; they make no statement about arbitrary physical noise.

\section*{Data and code availability}

The unchanged code-and-data archive \texttt{paperF\_code\_data\_v1.0.0.zip} accompanies this revision as an arXiv ancillary file. It contains the original frozen F1--F4 inputs, code, results and provenance; the F5-S synthetic protocol; the reference and budget analyses; the exact decoder, matrix, channel and distillation checks; and the v0.4 root-domain check. The archive includes an environment specification, a combined verification command, and SHA-256 integrity manifests. Its README distinguishes regeneration from checks of stored results and explains the stage-isolation requirement for historical upstream seals. Manuscript sources and historical article drafts are excluded from the ancillary archive.

The numerical interface values remain those of the frozen calculations. The v0.2 displayed decoder probability and classical/Pauli count wording were incorrect; the corrected formula is equation~\eqref{eq:decoderbound}, and the integer enumerator is unchanged. Historical correction and review records are retained without rewriting sealed campaign data. The November 13 release changed publication metadata and availability wording only. This revision adds higher-charge-superconductor context and references and clarifies the model scope; the mathematical results and numerical calculations are unchanged. It adds no physical realization or device-noise claim. The work has received computational and AI review but no independent human verification.

A DOI was reserved for the November 13 reproducibility archive, including that manuscript version and historical provenance: \href{https://doi.org/10.5281/zenodo.24185612}{10.5281/zenodo.24185612}. The code-and-data package is provided with this preprint as an ancillary file.

\setlength{\bibsep}{2pt}


\begin{thebibliography}{99}

\bibitem{Shi2026}

Zhengyan Darius Shi, Zhaoyu Han, Srinivas Raghu, and Ashvin Vishwanath. Charge-{$4e$} superconductor with parafermionic vortices: A path to universal topological quantum computation. (2026). \href{https://arxiv.org/abs/2602.06963v2}{arXiv:2602.06963v2}.

\bibitem{BravyiKitaev2005}

Sergey Bravyi, and Alexei Kitaev. Universal quantum computation with ideal {Clifford} gates and noisy ancillas. \emph{Physical Review A} \textbf{71}, 022316 (2005). \href{https://doi.org/10.1103/PhysRevA.71.022316}{doi:10.1103/PhysRevA.71.022316}.

\bibitem{Prakash2020}

Shiroman Prakash. Magic state distillation with the ternary {Golay} code. \emph{Proceedings of the Royal Society A} \textbf{476}, 20200187 (2020). \href{https://doi.org/10.1098/rspa.2020.0187}{doi:10.1098/rspa.2020.0187}.

\bibitem{CuiHongWang2014}

Shawn X. Cui, Seung-Moon Hong, and Zhenghan Wang. Universal quantum computation with weakly integral anyons. \emph{Quantum Information Processing} \textbf{14}, 2687--2727 (2015). \href{https://doi.org/10.1007/s11128-015-1016-y}{doi:10.1007/s11128-015-1016-y}.

\bibitem{Chen2025}

Liyuan Chen, Yuanjie Ren, Ruihua Fan, and Arthur Jaffe. A universal circuit set using the {$S_3$} quantum double. \emph{npj Quantum Information} \textbf{11}, 112 (2025). \href{https://doi.org/10.1038/s41534-025-01063-4}{doi:10.1038/s41534-025-01063-4}.

\bibitem{Gao2026Clustering}

Zhi-Qiang Gao, Yan-Qi Wang, Hui Yang, and Congjun Wu. Topological Charge-$2ne$ Superconductivity and Clustering Hierarchy. (2026). \href{https://arxiv.org/abs/2512.21325v5}{arXiv:2512.21325v5}.

\bibitem{Wang2026Charge6e}

Yan-Qi Wang, Boran Zhou, Hui Yang, and Zhi-Qiang Gao. Charge-$6e$ superconductivity from doping $SU(3)$ spin liquids. (2026). \href{https://arxiv.org/abs/2607.25909v1}{arXiv:2607.25909v1}.

\bibitem{Gao2026SCSharp}

Zhi-Qiang Gao, Yan-Qi Wang, Hui Yang, Zhaoyu Han, and Ashvin Vishwanath. SC$^\sharp$: superconductivity intertwined with topological order. (2026). \href{https://arxiv.org/abs/2610.10679v1}{arXiv:2610.10679v1}.

\bibitem{Bartlett2007}

Stephen D. Bartlett, Terry Rudolph, and Robert W. Spekkens. Reference frames, superselection rules, and quantum information. \emph{Reviews of Modern Physics} \textbf{79}, 555--609 (2007). \href{https://doi.org/10.1103/RevModPhys.79.555}{doi:10.1103/RevModPhys.79.555}.

\bibitem{Barkeshli2015}

Maissam Barkeshli, Parsa Bonderson, Meng Cheng, and Zhenghan Wang. Symmetry fractionalization, defects, and gauging of topological phases. \emph{Physical Review B} \textbf{100}, 115147 (2019). \href{https://doi.org/10.1103/PhysRevB.100.115147}{doi:10.1103/PhysRevB.100.115147}.

\bibitem{Hutter2015}

Adrian Hutter, and Daniel Loss. Quantum computing with parafermions. \emph{Physical Review B} \textbf{93}, 125105 (2016). \href{https://doi.org/10.1103/PhysRevB.93.125105}{doi:10.1103/PhysRevB.93.125105}.

\bibitem{Appleby2005}

D. M. Appleby. Symmetric informationally complete--positive operator valued measures and the extended {Clifford} group. \emph{Journal of Mathematical Physics} \textbf{46}, 052107 (2005). \href{https://doi.org/10.1063/1.1896384}{doi:10.1063/1.1896384}.

\bibitem{CuiWang2015}

Shawn X. Cui, and Zhenghan Wang. Universal quantum computation with metaplectic anyons. \emph{Journal of Mathematical Physics} \textbf{56}, 032202 (2015). \href{https://doi.org/10.1063/1.4914941}{doi:10.1063/1.4914941}.

\bibitem{Gross2006}

David Gross. {Hudson}'s theorem for finite-dimensional quantum systems. \emph{Journal of Mathematical Physics} \textbf{47}, 122107 (2006). \href{https://doi.org/10.1063/1.2393152}{doi:10.1063/1.2393152}.

\bibitem{Campbell2012}

Earl T. Campbell, Hussain Anwar, and Dan E. Browne. Magic-state distillation in all prime dimensions using quantum {Reed--Muller} codes. \emph{Physical Review X} \textbf{2}, 041021 (2012). \href{https://doi.org/10.1103/PhysRevX.2.041021}{doi:10.1103/PhysRevX.2.041021}.

\bibitem{Bonderson2008}

Parsa Bonderson, Kirill Shtengel, and J. K. Slingerland. Interferometry of non-{Abelian} anyons. \emph{Annals of Physics} \textbf{323}, 2709--2755 (2008). \href{https://doi.org/10.1016/j.aop.2008.01.012}{doi:10.1016/j.aop.2008.01.012}.

\end{thebibliography}
\end{document}